# Quasi-solid-state sodium-ion hybrid capacitors enabled by UiO-66@PVDF-HFP multifunctional separators: selective charge transfer and high safety

Wenliang Feng [a, b, 1], Jing Zhang [a, b, 1], Abdulmalik Yusuf [a, b], Xiang Ao [a, b], Dongfeng, Shi [a, b], Vinodkumar Etacheri [a]*, and De-Yi Wang [a]*

[a] IMDEA Materials Institute, C/ Eric Kandel 2, Getafe, Madrid 28906, Spain.

[b] Universidad Politécnica de Madrid, E.T.S. de Ingenieros de Caminos, 28040 Madrid, Spain.

[1] Wenliang Feng and Jing Zhang contributed equally to this work.

Corresponding Author:

E-mail address: vinodkumar.etacheri@imdea.org (V. Etacheri), deyi.wang@imdea.org (D.-Y. Wang).

**Abstract**

The practical application of sodium-ion hybrid capacitors is limited by their low energy densities resulted from the kinetics mismatch between cathodes and anodes, and the fire safety related to the flammable electrolyte-separator system. Hence, we report a rational design of metal-organic frameworks (MOFs, UiO-66) modified PVDF-HFP separator. High tensile strength and dimensional thermal stability of the separator reduce the risk of electrode short circuit caused by the separator deformation. MCC test demonstrates a reduction of 75% in peak heat release rate (pHRR), indicating an enhanced fire-resistant property of the separator. This is due to the transformation of UiO-66 into $ZrO_2$ accompanied by the consumption of oxygen and the formation of the barrier char that suppresses further heat release. Quasi-solid-state electrolyte prepared based on this separator presents an enhanced ionic conductivity of 2.44 mS·cm$^{-1}$ and Na-ion transference number of 0.55, which are related to the high porosity (> 70%) and electrolyte uptake (~ 320%) of the separator. Moreover, the open metal sites of UiO-66 can capture $PF_6^-$ and consequently liberate the $Na^+$ for faster migration, thus reducing the kinetics mismatch between cathodes and anodes. Such multifunctional separator enables the quasi-solid-state Na-ion hybrid capacitor to achieve high energy density (182 Wh·kg$^{-1}$ @31 W·kg$^{-1}$) and power density (5280 W·kg$^{-1}$ @22 Wh·kg$^{-1}$), as well as excellent cyclic stability (10000 cycles @1000 mA·g$^{-1}$).

**Keywords:** Quasi-solid-state; PVDF-HFP; Metal-organic frameworks; Dimensional thermal stability; Fire safety; Selective charge transfer

**1. Introduction**

Next-generation energy storage devices with large capacity, fast charging/discharging, long lifespan, low cost, and high safety have become imperative to meet the ever-increasing market demand for their widespread use [1-3]. Recently, sodium-ion hybrid capacitors (SHCs) have captured extensive attention due to the complementary advantages of both batteries (high energy density) and supercapacitors (high power density and cyclic stability) [4, 5]. Moreover, developing SHCs is more economically favorable, especially for large-scale applications benefiting from the wide availability of sodium resources [2]. The typical SHC is fabricated with a battery-type anode dominated by Na-ion diffusion behavior and a capacitor-type cathode controlled by physical adsorption-desorption behavior. The different storage mechanisms of individual electrodes lead to a kinetics imbalance between sluggish "Faradaic reaction" and fast "non-Faradaic reaction", which is the key reason for its poor rate and cycling performance [5-7]. Many efforts to address this issue are mainly focusing on exploring proper anode materials for fast Na-ion insertion/desertion. For example, insertion-type anodes based on $TiO_2$, $Na_2Ti_3O_7$, $NaTi_2(PO_4)_3$, $Nb_2O_5$, $V_2O_5$ or their carbon-based (graphene, rGO, CNT, etc.) hybrids, MXene ($Ti_2C$, $V_2C$), and biomass-derived carbons have been reported previously for SHCs due to their intrinsic or extrinsic pseudocapacitance and rapid Na-ion intercalation [7-18]. However, the accelerated Na-ion insertion/desertion is ineluctably at the expense of capacity due to the much more sluggish Na-ion diffusion compared to the fast absorption of anions in nature. The electrode polarization caused by the unbalanced charge numbers can further restrict the charge separation and Na-ions migration, and thus results in severe capacity fading and inadequate energy density. Relying solely on optimization of anodes could not realize comprehensively high-performance SHCs. Besides, the serious safety issues related to flammable separators, electrolyte leakage, and ignition impede their large-scale applications [1, 19-21], which exist in all types of energy storage devices based on organic electrolytes. Although considerable non-flammability was achieved through the addition of flame retardants into the electrolytes, the decreased ionic conductivity could significantly deteriorate the electrochemical performance [20-22]. Electrode short circuits related to separator deformations (puncture or thermal shrinkage) also give rise to a catastrophic failure of these devices [23].

As an effective strategy, modifications of the separator-electrolyte system, the bridge for charge shuttling, and protection wall for battery operation have demonstrated great potential for fast ion kinetics, fire safety, electrolyte saving, and leakage proof. Functional coating materials (Polymers, MXenes, Nanofibers, $Al_2O_3$, $Al(OH)_3$, $SiO_2$, $TiO_2$, and $ZrO_2$ nanoparticles, etc.) have been adopted to modify the polyethylene (PE) or polypropylene (PP) membranes, the most commonly used separators in commercial Li-ion batteries [24-28]. These bilayer separators exhibit good mechanical strength and improved thermal stability from the coating materials. However, these efforts achieved only marginal improvement due to the inherent drawbacks of PE and PP, such as deficient thermal stability, poor wettability to electrolytes, and low porosity (~ 40%), resulting in high internal resistance for ion transport, especially for Na-ions with larger radius (1.02 Å) compared to Li-ions (0.76 Å). Moreover, the issues related to electrolyte leakage remain unsolved.

Recently, poly(vinylidene fluoride-hexafluoropropylene) (PVDF-HFP) based membranes with multi-scale pore structures have been widely studied and demonstrated many advantages as separators for energy storage devices [29-32]. (i) PVDF-HFP has good chemical stability and wettability toward common organic solutions to form quasi-solid-state (QSS) electrolytes thus preventing electrolyte leakage [33]. (ii) PVDF-HFP with a relatively low melting point (~ 140 °C) ensures that the pores of separators close at the early stage of combustion and shut down the battery before the catastrophic failure happens [34]. The $β$-phase polymer chains of PVDF-HFP are electroactive and beneficial for the charge transfer in separators [35]. (iii) PVDF-HFP can be easily processed into membranes with tunable porous structures at a micrometer scale [36]. (iv) PVDF-HFP based membranes have excellent mechanical strength, thermal stability, and non-flammability [23, 36]. In addition, the incorporation of active fillers into the PVDF-HFP polymer matrix has presented further improvement in ionic conductivity, mechanical strength, or dimensional thermal stability [23, 36-39]. Benefiting from the inorganic-organic hybrid nature, metal-organic frameworks (MOFs) offer multiple advantages as active fillers of good compatibility with polymer-based separators. The pores of MOFs consisting of angstrom-sized windows and nanometer-sized cavities have been reported to work as selective filters and conducting channels for alkali metal ions [40, 41]. Their open metal sites could potentially coordinate with anions to liberate charge mobility [19, 42, 43]. Mesoporous structure of MOFs with high surface and porosity can act as a liquid electrolyte reservoir to improve the wettability of separators and thus reduce interfacial resistance [36]. Moreover, MOFs and their derivatives possess abundant fire-retardant elements, potential carbon sources, and transition metal species that exhibit great improvement in the thermal stability and flame retardancy of polymer substrates [44, 45]. Transition metal zirconium has been proved as an effective fire-retardant element, for example α-zirconium phosphate ($α$-ZrP). This resulted from the unsaturated metal and formation of $ZrO_2$ that promote the catalytic carbonization process [46, 47]. It facilitates the formation of a barrier and thus protects the underlying polymer substrate. Therefore, Zr-based MOFs can be potential candidates as multifunctional additives to comprehensively improve the performance of PVDF-HFP separators. However, to our knowledge, there are very few or no works reported to exploit these unique properties of PVDF-HFP and MOFs especially for Na-ion based energy storage systems.

Herein, we propose a high-performance quasi-solid-state Na-ion hybrid capacitor enabled by a multifunctional UiO-66/PVDF-HFP hybrid separator. Activated UiO-66 is selected as active fillers into the PVDF-HFP based on the following aspects.[36, 48, 49] (i) Their fully open pores and metal sites can absorb more electrolyte solution and selectively transport Na-ions. The acid defect positions of UiO-66 have been proved highly beneficial for ion transport, thus reducing the kinetics mismatch between cathodes and anodes. (ii) The function as a flame retardant additive ensures better operating safety of Na-ion hybrid capacitor. (iii) UiO-66 has good chemical stability in organic electrolytes that avoids the deterioration and contamination of electrolytes caused by side reactions. (iv) UiO-66 is unreactive to Na-ions that can avoid Na dendrites caused by the Na-ion deposition on the MOF/separator and thus reduce the risk of internal short circuit. (v) The fast and low-cost synthesis of UiO-66 makes the separators more economically favorable.

## 2. Experimental section

*2.1 Chemicals and materials*

Zirconium (IV) chloride (ZrCl$_4$, ≥ 99.8%, Alfa Aesar), Terephthalic acid (H$_2$BDC, ≥ 98.0%, Sigma-Aldrich), Hydrochloric acid (HCl, ≥ 37%, Sigma-Aldrich), N,N-dimethylformamide (DMF, ≥ 99.8%, Sigma-Aldrich), Poly(vinylidene fluoride-co-hexafluoropropylene) (PVDF-HFP, battery grade, Kynar-Flex®2801, Arkema Inc.), Sodium hexafluorophosphate (NaPF$_6$, 99.99%, Alfa Aesar), propylene carbonate (PC, 99.99%, Sigma-Aldrich), Titanium(III) chloride (TiCl$_3$, 99.9%, Acros Organics), ethylene glycol (EG, 99.9%, Fisher Scientific), polyvinylidene fluoride (PVDF, MW: 600,000, MTI Chemicals), N-Methyl-2-Pyrrolidone (NMP, 99.9%, Aladdin Chemicals), Celgard (MTI), glass fiber (GF/B, Whatman) were used without any further treatment.

*2.2 Synthesis of metal-organic framework, UiO-66*

The zirconium terephthalate framework (UiO-66) was prepared *via* a solvothermal reaction followed by a heat treatment under an inert atmosphere [50]. Briefly, 0.85 g of ZrCl$_4$, 0.82 g of H$_2$BDC, and 4.0 mL of HCl were dissolved in 50 mL of DMF. The homogeneous solution was then loaded into a Teflon autoclave reactor and kept at 120 °C for 24 h. The precipitate was collected through centrifugation and thoroughly washed with methanol followed by drying at 80 °C under vacuum for 12 h. The white powder obtained was calcined at 400 °C for 1 h under Ar-flow to get the resulting product.

*2.3 Preparation of UiO-66/PVDF-HFP hybrid separators and quasi-solid-state electrolytes*

The UiO-66/PVDF-HFP hybrid membranes were prepared drawing on the experience of previous reports with minor modification [51, 52]. Briefly, 0.7 g of PVDF-HFP was dissolved into 4.0 g of DMF and acetone mixture (weight ratio of 7:1) under stirring for 1h at 25 °C to get a transparent solution. The UiO-66 powder with a specific weight was dispersed into the above solution under ultrasonication for 30 min and stirred for 2 h successively. The homogeneously mixed solutions obtained with different UiO-66 concentrations, 0 wt%, 3 wt%, 6 wt%, and 9 wt%, are defined as PVDF-HFP, 3-Hybrid, 6-Hybrid, and 9-Hybrid, respectively. The solutions were then individually coated on a glass substrate using a doctor blade with a gap size of 400 μm. The membranes obtained were immersed into a 9:1 mixture solution of water and ethanol at 25°C for 12 h and then transferred into fresh water solution to remove the track solvent. After drying at 80 °C for 6 h under vacuum, the membranes were cut into circular disks with a diameter of ~1.6 cm using as separators. The separators obtained were then immersed into the liquid electrolyte solution (1.0 M $NaPF_6$ in PC) for 5 h to form quasi-solid-state (QSS) electrolytes for the subsequent experiments.

*2.4. Characterization techniques*

Field emission scanning electron microscopy (FE-SEM) images were collected using an FEI Helios NanoLab 600i microscope equipped with energy-dispersive X-ray spectroscopy (EDS) at the voltage of 5.0 kV. X-ray diffraction (XRD) patterns were recorded using a PANalytical Empyrean with Cu *Kα* radiation (*λ*=1.54 Å). Thermogravimetric analysis (TGA) was performed on TA Q50 with a heating rate of 10 °C·$min^{-1}$ under the nitrogen atmosphere. Fourier transform infrared spectroscopy (FTIR) was performed using a Nicolet iS50 spectrometer. Differential scanning calorimetry (DSC) was recorded using Q200, TA Instruments at a heating rate of 10 °C·$min^{-1}$. The fire behavior of all separators was investigated using Micro-scale Combustion Calorimeter (MCC, FTT) with a sample weight of ~ 5.0 mg. The system was heated to 700 °C at a heating rate of 1.0 °C·$s^{-1}$ under a 20 $cm^3$·$min^{-1}$ gas stream consisting of oxygen (20%) and nitrogen (80%).

*2.4.1. Determination of thermal shrinkage, electrolyte uptake, ionic conductivity, MacMullin number, and charge transference number*

The dimensional thermal stability of the separators is evaluated by analyzing their thermal shrinkage ratio after heating at a specified temperature for 10 min. Thermal shrinkage ratio ($T_s$, %) is calculated by the equation:

$$T_s = \frac{S_0}{S} \times 100\% \quad (1)$$

where $S_0$ and $S$ are the surface area of the separators before and after thermal treatment, respectively.

The circular separators were pre-soaked in liquid electrolyte (1.0 M NaPF$_6$ in PC) for 5 h. Then extra liquid electrolyte on the surface of the separators was carefully wiped with a filter paper. The electrolyte uptake ($U$, %) was calculated according to the equation:

$$U = \frac{W - W_0}{W_0} \times 100\% \quad (2)$$

where $W_0$ and $W$ are the weights of the separators before and after soaking in liquid electrolyte, respectively.

The ionic conductivity of the QSS electrolytes was measured in the temperature range of 25−85 °C on the stainless-steel|electrolyte|stainless-steel cells. It was done using an electrochemical workstation (ZIVE-SP1) in the frequency range of 0.1−1×10$^6$ Hz with an AC amplitude of 5.0 mV and calculated according to the equation:

$$\sigma = \frac{l}{R \times S} \quad (3)$$

where $\sigma$ (mS·cm$^{-1}$) is the ionic conductivity, $l$ (μm) is the thickness; $S$ (cm$^2$) is the effective area; $R$ (Ω) is the bulk resistance of the QSS electrolytes.

The relationship between ionic conductivity (σ) and temperature ($T$) can be described by the Arrhenius equation:

$$\sigma(T) = A exp\left(\frac{E_a}{RT}\right) \quad (4)$$

where σ, $A$, $R$, and $E_a$ are the specific conductivity, pre-exponential factor, universal gas constant (8.314 J·mol$^{-1}$), and the activation energy, respectively. where σ is the conductivity,

The MacMullin number ($N_m$) was calculated using the equation:

$$N_m = \frac{\sigma_0}{\sigma} \quad (5)$$

where $\sigma_0$ and σ are the ionic conductivities of the liquid electrolyte and the QSS electrolytes at 25 °C, respectively.

Na-ion transference number ($t_{Na+}$) of the QSS electrolytes was determined by AC impedance and DC potentiostatic polarization performed on Na|electrolyte|Na cells using ZIVE-SP1 workstation following Bruce-Vincent method:

$$t_{Na+} = \frac{I_{ss}(V - I_0 R_0)}{I_0(V - I_{ss} R_{ss})} \quad (6)$$

where $I_0$ and $R_0$ are initial current and interfacial resistance; $I_{ss}$ and $R_{ss}$ are the steady-state current and interfacial resistance; $V$ is the applied DC potential of 10 mV. The interfacial impedance was obtained from the Nyquist plots in the frequency range of $0.1-1\times10^6$ Hz.

*2.4.2. Electrochemical performance testing*

Dual-phase TiO$_2$ nanosheet anode, activated carbon (AC) cathode, and NaPF$_6$/PC electrolyte were prepared according to the previous report (see details in SI) [7]. Na-ion half-cells were assembled with TiO$_2$ nanosheet anodes, Na-foil counter electrodes, and QSS electrolytes (PVDF-HFP, 3-Hybrid, 6-Hybrid, and 9-Hybrid) in an Ar-filled glovebox (Vigor Tech, O$_2$ and H$_2$O levels < 0.1 ppm). QSS Na-ion hybrid capacitor was fabricated using the same method by replacing Na-foil with AC cathode. Individual electrodes were pre-cycled in their corresponding voltage ranges to reduce the irreversible capacity loss related to the formation of the solid electrolyte interface (SEI). Polypropylene (Celgard), the most commonly used commercial separator, was also adopted for a parallel comparison of their electrochemical performances. Galvanostatic charge-discharge tests were performed using a battery tester (Neware BTS-4000). Cyclic voltammetry (CV) and Electrochemical impedance spectroscopy (EIS) was performed using a ZIVE-SP1 an electrochemical workstation (see details in SI).

**3. Result and discussion**

*3.1 Synthesis and characterization of UiO-66/PVDF-HFP hybrid membrane*

The UiO-66 was prepared *via* a typical solvothermal reaction followed by calcination under an inert atmosphere (**Fig. 1, step 1**). The high-temperature activation process was done to remove the moisture, solvent, and residual reactants from the pores, and thus ensure high purity of UiO-66 with relatively full open pores and metal sites [40, 53]. The removal of guest molecules out of the organic frameworks could lead to a volume contraction of the unit cells possibly due to the strong edge-to-face π-π stacking interactions between the aromatic rings of the organic linkers [54]. Thermogravimetric analysis (TGA) was initially performed to evaluate the thermal stability of as-prepared UiO-66 (**Fig. S1a**). The first weight loss under 100 °C represents the desorption of physisorbed moisture, while the second weight loss between 100 and 450 °C is related to the removal of residual DMF and the dihydroxylation of the zirconium-oxo clusters [55]. The rapid weight loss after 450 °C is due to the decomposition of the organic linkers in the UiO-66 frameworks. This result suggests good thermal stability of UiO-66 below 400 °C. To avoid the structural transformation/decomposition, the heating temperature of 400 °C was determined to achieve a complete activation of UiO-66. Unchanged crystal structure of UiO-66 before and after high-temperature activation can be proved by the XRD result (**Fig. S1b**). The PVDF-HFP and UiO-66/PVDF-HFP hybrid membranes were prepared through a non-solvent induced phase separation (NIPS) process (**Fig. 1, step 2**) [51, 52]. This method is adopted due to its simple tunability of porosity degree and pore sizes of membranes. Asymmetric porous morphology of the membranes is realized through the diffusion exchange of the solvent (DMF and acetone) in PVDF-HFP solution with the non-solvent (water and ethanol) in a coagulation bath at 25 °C. The properties of the porous structure can be strongly affected by the exchange rate between the solvent and non-solvent [56, 57]. In principle, a higher exchange rate leads to larger pore size and uneven pore distribution. In this work, the exchange rate was controlled by adjusting the additive amount of ethanol in water solution as the PVDF-HFP/DMF/acetone system has stronger thermodynamic stability in ethanol compared to water solution.

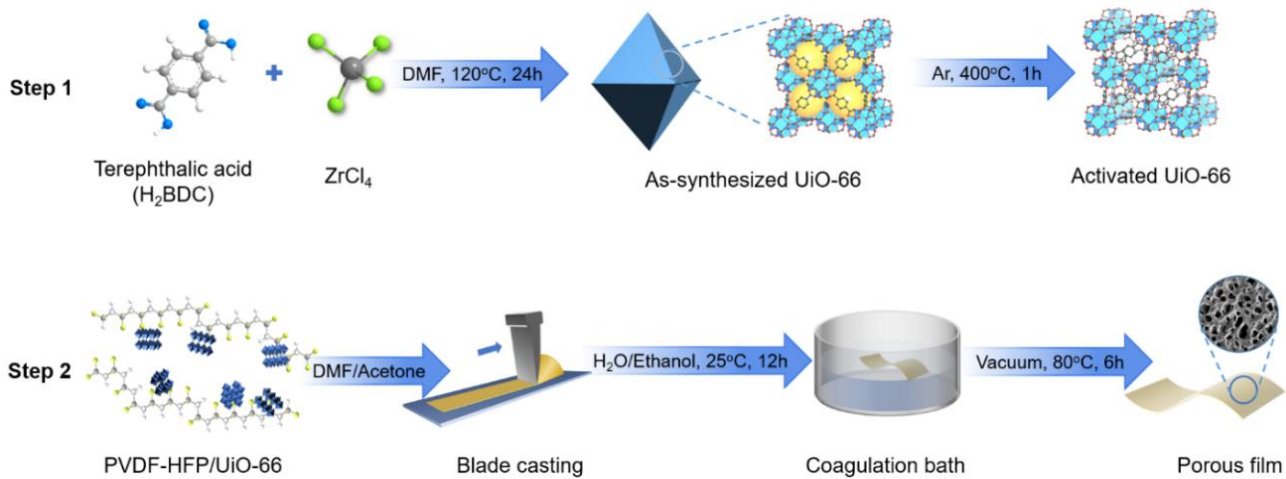

**Fig. 1.** Schematic illustration of the UiO-66 synthesis and the hybrid separator preparation.

Detailed microstructure analysis of the membranes was performed using scanning electron microscopy (SEM). The PVDF-HFP membrane (**Fig. 2a-c**) exhibits an asymmetric porous structure with a pore size of 1–2 μm and thickness of ~ 60 μm, which dramatically differ from the commercial Celgard (**Fig. S2a**) with a pore size of 200–300 nm, thickness of ~ 25 μm and porosity of ~ 40%. The larger pore size and higher porosity of the PVDF-HFP membrane observed here are beneficial for better electrolyte wettability and higher electrolyte uptake. It consequently reduces charge transfer resistance, ensures superior electrolyte/electrode contact, and provides sufficient Na-ions for quasi-solid-state SHCs. Notably, the incorporation of UiO-66 and PVDF-HFP further enlarged the pore size (1–3 μm) and porosity (> 70%) of hybrid membranes (**Fig. 2d-f, Fig. S2b, and S2c**). Such morphology change is possibly due to the hydrophilicity of UiO-66 that accelerates the exchange rate of solvent and non-solvent during the pore formation process. However, overloading of UiO-66 resulted in a severe non-uniform distribution of pores, uncontrollable increase of pore size, and pore closure (**Fig. S2d**). The oversize pores could increase the risk of internal short circuit caused by metal dendrites or the direct contact of electrodes. Moreover, the non-uniform distribution of pores could limit the utilization of active materials on electrodes, reducing the energy density and accelerating the degradation of SHCs. Uniform distribution of F, C, O, and Zr elements are evident from the EDX elemental mapping (**Fig. 2g**) of the hybrid membrane, indicating good miscibility of UiO-66 and the PVDF-HFP substrate. Electrolyte wettability was observed by dropping electrolyte solution on the surface of separators and photos were recorded after 60 s (**Fig. S3a**). In comparison to the Celgard, the PVDF-HFP and hybrid separators demonstrate much better wettability due to the good affinity of PVDF-HFP and UiO-66 fillers toward PC electrolyte and their high porosity that provides large space for electrolyte adsorption. Electrolyte uptake of the separators was calculated and listed in **Table S1**. Higher electrolyte uptake capability of PVDF-HFP separator (~ 250%) compared to commercial Celgard (~ 80%) supports that PVDF-HFP absorbs plenty of electrolyte solution to form gel electrolyte (**Fig. S3b**). This is essential for the fabrication of QSS SHCs with sufficient Na-ion supply and excellent electrolyte retention. Hybrid separators present further improved electrolyte uptake up to 320% resulting from the introduction of activated UiO-66 with fully open pores that is beneficial for trapping more electrolyte solution.

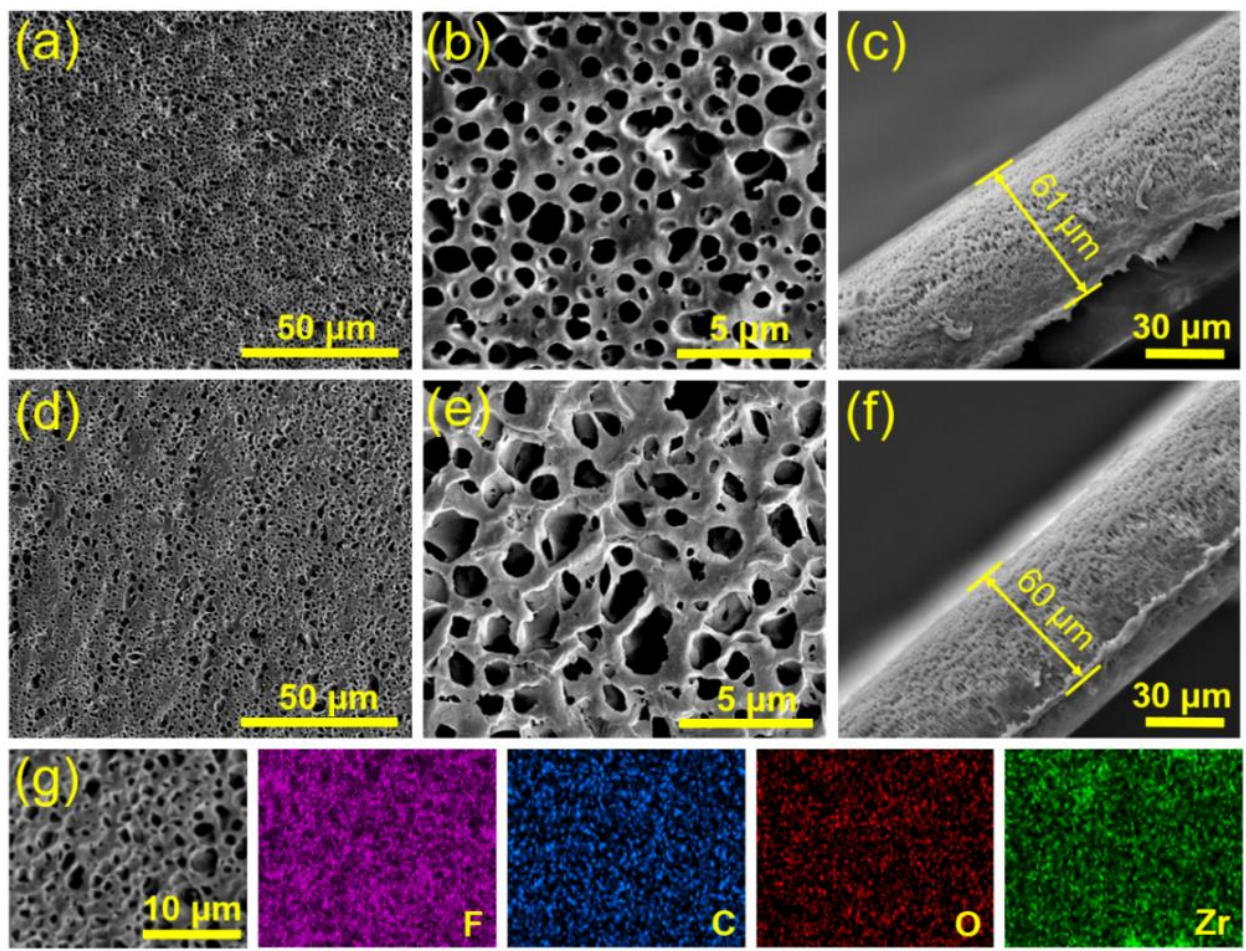

**Fig. 2.** Surface and cross-section SEM images of (a-c) PVDF-HFP, and (d-f) hybrid membrane. (g) HAADF image and corresponding EDX elemental mapping of the hybrid membrane.

The high mechanical strength of separators is essential for the fabrication of SHCs to ensure their safe operation. Catastrophic failure of these devices often follows from the electrode short circuit related to the separator deformation (puncture or thermal shrinkage) [58]. A tensile test of the membranes was carried out to characterize their mechanical properties. The hybrid membrane (**Fig. 3a**) demonstrates improved tensile strength (6.62 MPa) and elongation (107%) compared to the PVDF-HFP separator (5.31 MPa and 61%), which is very close to that of the commercial Celgard (8.96 MPa and 103%). The increased elongation of the hybrid separator suggests the good flexibility that is beneficial for a superior electrode/electrolyte contact, thus reducing the charge transfer resistance. The mechanical robustness of these hybrid membranes increases with an increase of UiO-66 loading within a certain range (**Fig. S4a**). Although the high porosity of membranes is usually believed to reduce their mechanical strength, the developed hybrid membranes exhibit both high porosity and satisfactory mechanical properties for practical applications. The superior mechanical properties of hybrid membranes compared to pure PVDF-HFP membrane are credited to the interfacial adhesion of UiO-66 fillers and PVDF-HFP matrix. In this case, UiO-66 provides numerous contact points to PVDF-HFP chains to enhance the interaction force between polymer chains, thus protecting the hybrid separator from easy breaking [23, 39].

The X-ray diffraction (XRD) analysis (**Fig. S4b**) reveals that the PVDF-HPF powder purchased is $α$-phase dominated with some traces of $β$-phase coexisting. This can be evidenced by the strong diffraction peaks at 18.4°, 19.9°, 26.7° and 38.7° corresponding to (020), (110), (021) and (200) of $α$-phase crystal and the weak peak at 35.9 corresponding to (200) of $β$-phase crystal, respectively [35, 59]. While the PVDF-HFP membrane prepared with a weaker peak at 18.4° and an intensive peak at 20.3° related to (020) and (200,110) are dominated in $β$-phase. With the increase of the UiO-66 loading in these hybrid membranes, the peak area at 20.3° decreases (**Fig. 3b and Fig. S4c**), indicating a reduced crystallinity of PVDF-HFP. The improved motion of amorphous PVDF-HFP polymer segments is thus expected to further promote the migration of Na-ions in the QSS electrolytes [35]. The corresponding Fourier transform infrared (FTIR) spectra (**Fig. S4d, S4e, and Fig. 3c**) are in good agreement with the XRD result. The specific bands of PVDF-HFP powder at 761, 793, 874, 971, 1182, 1401 cm$^{-1}$ indicate a dominated $α$-phase coexisting with $β$-phase traces (small peaks at 840, 1068, and 1282 cm$^{-1}$) [59]. The peak of $β$-phase at 840 cm$^{-1}$ enhanced in PVDF-HFP based membranes, while the peaks of $α$-phase disappeared or weakened, indicating a transformation of $α$-phase to $β$-phase. New peaks of UiO-66 can be also detected on the spectra of hybrid membranes due to the incorporating of UiO-66 and PVDF-HFP. In principle, the PVDF-HFP polymer with higher crystallinity would have a higher melting temperature ($T_m$) and heat of fusion ($\Delta H_f$) [60]. Differential scanning calorimetry (DSC) curves of the membranes (**Fig. 3d and Fig. S4f**) exhibit a decreased $T_m$ and $\Delta H_f$ with adding UiO-66 into the polymer matrix of PVDF-HFP. The reduced crystallinity of hybrid membrane demonstrated here is also well consistent with the conclusion of XRD results.

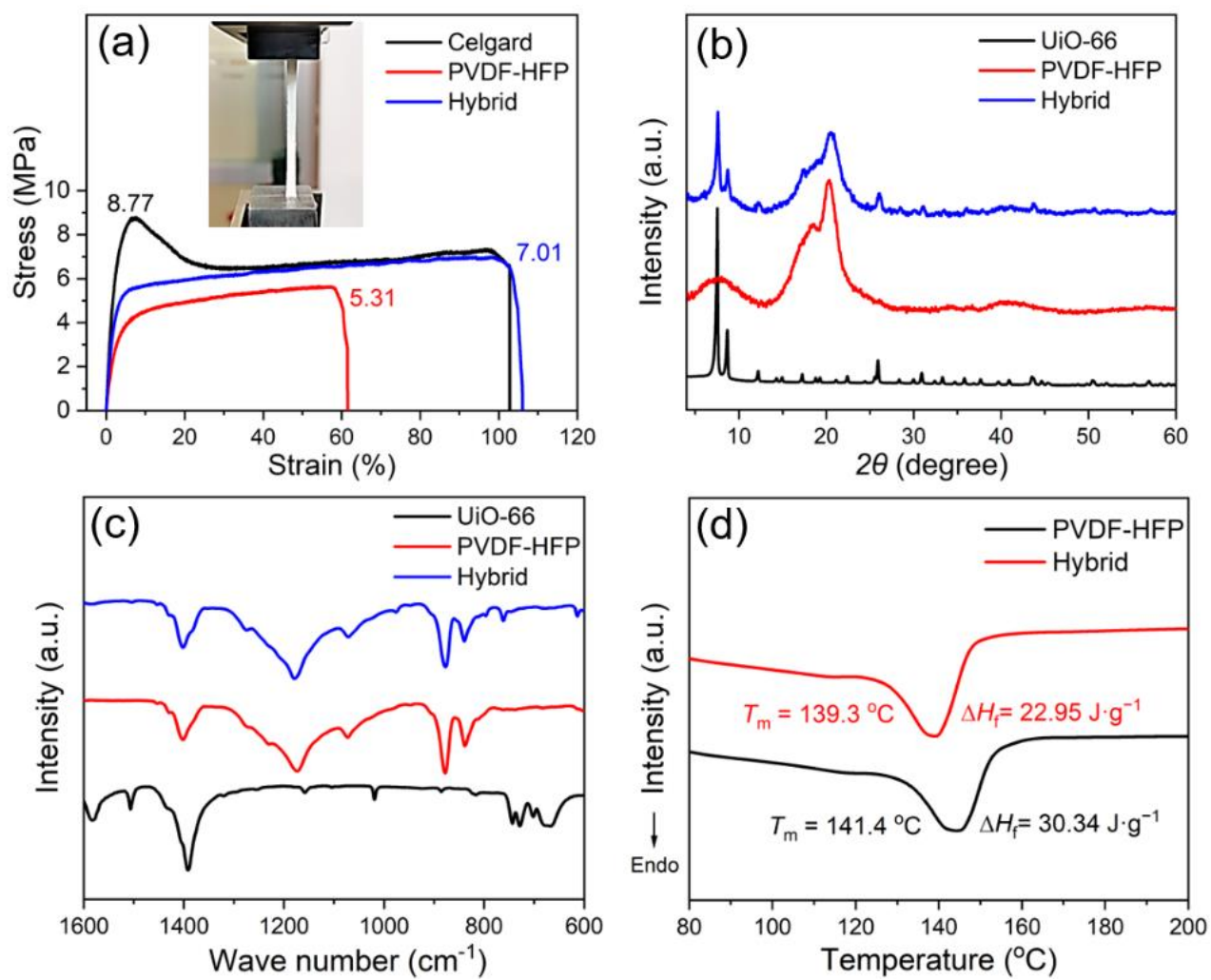

**Fig. 3.** (a) Stress-strain curves of Celgard, PVDF-HFP, and hybrid membranes. (b) XRD patterns and (c) FTIR spectra of the activated UiO-66, PVDF-HFP, and hybrid membranes. (d) DSC curves of the PVDF-HFP and hybrid membranes.

*3.2 Dimensional thermal stability and flame resistance of UiO-66/PVDF-HFP hybrid separators*

A minimal thermal shrinkage of separators is desperately required to prevent internal short circuit of SHCs at elevated temperatures. As presented in **Fig. 4a, Fig. S5a, and Table S1**, commercial Celgard starts to shrink rapidly at 140 °C and melts completely into a trunk-like residue at 160 °C. Such high thermal shrinkage of the Celgard could result in an increased safety risk to SHCs during thermal runaway. PVDF-HFP separator intrinsically has better thermal resistance capability due to its molecular properties. Only crinkles are observed around the edge of the PVDF-HFP separator at 140 °C, which could be avoided by proper pressure from both electrodes. However, more than 20% of thermal shrinkage occurring at 160 °C still cannot meet the requirement of SHC safety. In contrast, the hybrid separators exhibit invisible thermal shrinkage (< 7%) under identical conditions indicating that the UiO-66 can help to improve the dimensional thermal stability of hybrid separators. Notably, the pores of the hybrid separator fully closed after heating at 140 °C (**Fig. 4b**), which can effectively shut down SHCs against thermal runaway during an external short circuit, overcharge, and/or abusive discharge conditions [34]. TGA was also carried out to study the decomposition process of the separators under high temperatures (**Fig. 4c and Fig. S5b**). The commercial Celgard decomposes at 350 °C without any residue formed after 450 °C. The PVDF-HFP separator shows a much higher decomposition temperature of 400 °C and residue of 20% at 600 °C. By contrast, the residue of the hybrid separator greatly increases up to 40%, which is almost double that of PVDF-HFP. However, the slightly reduced decomposition temperature of the hybrid separators indicates that the UiO-66 can accelerate the decomposition of PVDF-HFP due to the "catalytic oxidation effect" [61, 62]. The halogen-based radicals released during the decomposition of PVDF-HFP can improve the fire safety of separators by diluting the flammable gases [63, 64]. Meanwhile, UiO-66 catalytically promotes the carbonization of PVDF-HFP. Such carbonaceous layer could also reduce the penetration of heat and oxygen supply through the sample, and accordingly prevent further exposure under fire.

To further study the fire behavior of the separators, the Micro-scale Combustion Calorimetry (MCC) test was employed to measure the key fire safety parameters such as heat release rate (HRR), heat release capacity (HRC), etc. **(Fig. 4d and Fig. S5c)**. The peak heat release rate (pHRR) of the Celgard is as high as 1126 W·g$^{-1}$. The PVDF-HFP separator demonstrates a much lower pHRR of ~ 700 W·g$^{-1}$ indicating the reduced danger during fire hazards compared to the commercial Celgard. Moreover, with the addition of 3 wt%, 6 wt%, and 9 wt% of UiO-66 into the hybrid separators, the pHRR values are found to be 263, 172, and 184 W·g$^{-1}$, exhibiting 62%, 75%, and 73% of reduction, respectively, compared to that of PVDF-HFP. The significantly suppressed heat generation indicates an improved fire safety of the separators. Interestingly, different from the single peak of the PVDF-HFP separator at 490 °C, the hybrid separators show two peaks at 417 °C and 467 °C, respectively. Moreover, the peak value of the first peak is proportional to the UiO-66 loading in the hybrid separator. The appearance of the first peak can be attributed to the presence of activated UiO-66, containing abundant zirconium nodes, that being oxidized into metal oxide. The $ZrO_2$ formation can be observed from the XRD spectra of the residue of the hybrid separator (**Fig. S5d**). Additionally, transition metal-containing UiO-66 can facilitate the formation of the char layer through the catalytic carbonization behavior, which is also consistent with the increased residue of hybrid separators compared to PVDF-HFP as presented in TGA. It is well established that the fire triangle contains three basic elements, heat, fuel, and oxygen, in the combustion process [1]. Therefore, the initial consumption of oxygen caused by UiO-66 reduced the ambient oxygen content, thus suppressing the heat releasing from PVDF-HFP (the second peak at HRR curve). To verify it, MCC measurement of a parallel hybrid separator containing UiO-66 derived $ZrO_2$, obtained by heating as-prepared UiO-66 powder at 500 °C under Ar-flow, was recorded under the same condition (**Fig. S5e**). The disappearance of the first peak at 417 °C qualitatively proves that the transformation of UiO-66 into $ZrO_2$ accompanied by the consumption of oxygen is responsible for the formation of the first peak. However, the total heat release (**Fig. S5f**) of all hybrid membranes is nearly constant with increasing the UiO-66 loading. Therefore, the overloading of UiO-66 will not further improve the fire safety of the separators. A small additive amount of UiO-66 is more economical for practical application.

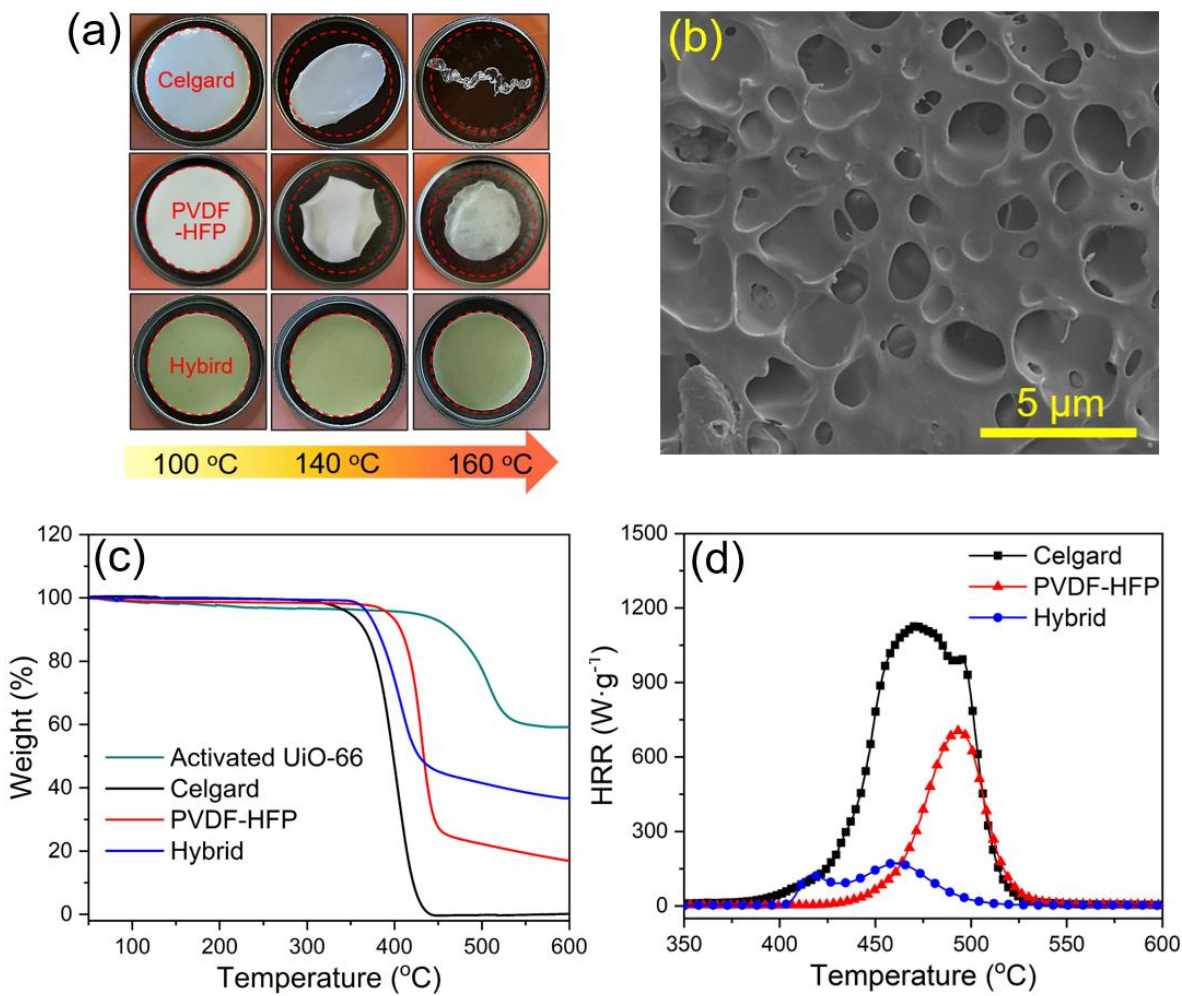

**Fig. 4**. (a)Thermal shrinkage test of the Celgard, PVDF-HFP, and hybrid separators. (b) SEM image of the hybrid separator after heating at 140 °C for 10 min. (c) TGA curves, and (d) HRR curves of the Celgard, PVDF-HFP, and hybrid separators.

The fire test for the separators upon a forced small flame was also performed. Each sample was exposed to a flame for 5 s and the photos during the burning test and final residues were recorded (**Fig. 5 and Fig. S6**). The Celgard separator shrinks immediately upon contact with the flame without any formation of residue, which is consistent with the TGA result. The shrinkage of the PVDF-HFP membrane is reduced under the same condition with the formation of a carbonaceous residue. Moreover, the hybrid separator exhibits superior thermal stability and maintains its shape, and the residue obtained after 5 s also shows the highest value. As observed from the SEM images (**Fig. S7**), the residue of the hybrid separator after the fire test presents a more intumescent char layer and smaller pore size compared to the common fragile residues [65, 66]. Moreover, the structure of the cross-section area is more continuous and compact. Such carbonaceous shield that initially formed on the surface of the separators can act as an insulating barrier and prevent the oxygen and heat from reaching deeper [65]. It is thus clear that incorporation of the UiO-66 and PVDF-HFP endows hybrid separators with high porosity, excellent mechanical robustness, good dimensional thermal stability, and flame-retardancy, and thus demonstrates great potential for SHCs applications.

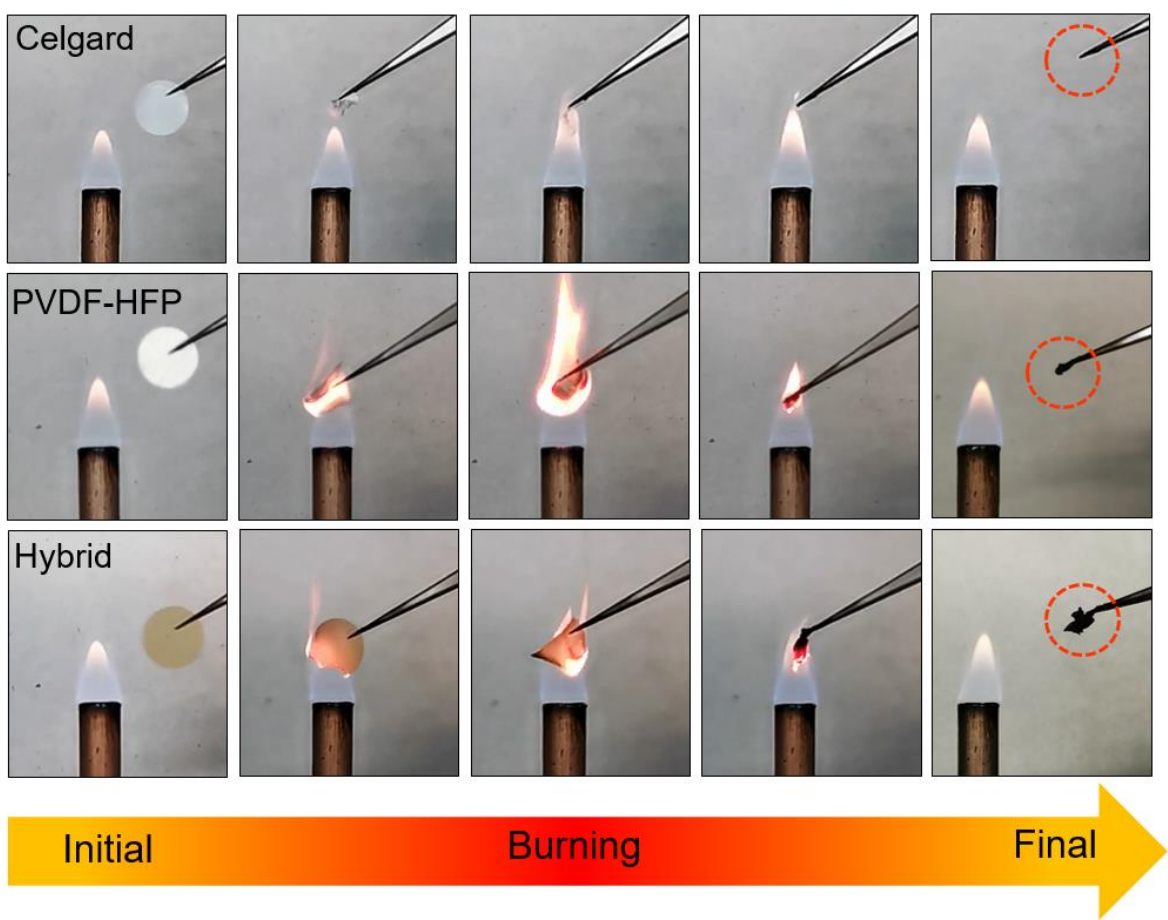

**Fig. 5.** Fire test of Celgard, PVDF-HFP, and hybrid separators

*3.3. Electrochemical performance of quasi-solid-state hybrid electrolyte*

The ionic-conductivity analysis of the quasi-solid-state (QSS) electrolytes based on different separators was performed using electrochemical impedance spectroscopy (EIS) in the temperature range of 25–85 °C (**Fig. 6a and Fig. S8a-e**). The corresponding resistances can be obtained from the Nyquist plots, where the AC responses present an inclined straight-line (the ion-blocking electrode behavior). The interception on the real axes, $Z'_{RE}$, provides the bulk ionic resistance of the QSS electrolytes [36, 49]. The ionic-conductivity values at 25 °C calculated by equation 3 are listed in **Table S2**. The hybrid electrolytes demonstrate superior ionic-conductivity (2.15–2.83 mS·cm$^{-1}$) compared to the PVDF-HFP (1.42 mS·cm$^{-1}$), which is ~6 times higher than the Celgard (0.38 mS·cm$^{-1}$). The conductivity comparison of the hybrid separator with previous work is listed in **Table S3**. The positive correlation between ionic-conductivity and UiO-66 loading of the hybrid electrolytes is independent of their electrolyte uptakes, which means that the improved ionic conductivity, in this case, is mainly resulting from the selective Na-ion transport supported by the UiO-66 filler and the increased motion of the polymer segments caused by the reduced crystallinity of the PVDF-HFP.

The temperature-dependent ionic conductivity of the QSS electrolytes obeys the Arrhenius type thermally activated process (equation 4) in the temperature range of 25–85 °C ($T < T_m$). The increased ionic conductivity toward higher temperature (**Fig. 6b and Fig. S8f**) can be explained by the free-volume theory of polymer [67]. The polymer matrix of the QSS electrolytes expands with an increase of temperature, thus providing a larger free volume to promote the motion of the polymer segments and the mobility of Na-ions. The activation energy ($E_a$) of the Na-ion migration in the QSS electrolytes was calculated using the slopes of ln ($\sigma$) *vs*. $T^{-1}$ (**Table S2**). The lower $E_a$ values of the QSS hybrid electrolytes (1.31–1.58 kJ·mol$^{-1}$) compared to that of the PVDF-HFP (2.03 kJ·mol$^{-1}$) and Celgard (2.63 kJ·mol$^{-1}$) indicate a lower energy barrier for Na-ion migration, which is essential for the fabrication of polymer-based QSS electrolytes [49]. It has been established that the liquid electrolyte entrapped in the pores of the polymer matrix will further swell the amorphous domains to form a gel state [68]. Thus, Na-ions can be transferred in porous polymer-based QSS electrolytes: (a) through the liquid solution absorbed in pores; (b) through swelled amorphous domains; and (c) along molecular chains in the polymer matrix. Since the Na-ion transfer process (c) is much more sluggish than process (a) and (b), the increased ionic conductivity of the hybrid QSS electrolytes compared to the PVDF-HFP and Celgard is mainly attributed to the higher liquid electrolyte absorption and amorphous domains in the polymer matrix. This is well consistent with the high porosity and electrolyte uptake, and the relatively low crystallinity of the hybrid membranes as discussed earlier. MacMullin number ($N_m$) was also calculated by equation 5 to provide the information regarding the ionic resistance related to the porous structure and the affinity between separator and electrolyte (**Table S2**) [69]. The lower $N_m$ values of the hybrid (3.62–2.75) and PVDF-HFP (5.49) separators suggest much lower ionic resistance and higher permeability of Na-ions compared to the Celgard (20.13) due to their high porosity and good affinity with electrolyte, which has been confirmed by the SEM and wettability observations.

Low Na-ion diffusivity in QSS electrolytes could lead to a serious concentration polarization of SHCs during the cycling process. It consequently results in poor charge separation, increased side reactions, and joule heating, thus reducing the rate performance and cycling stability of the devices, especially under fast charging-discharging conditions. Hence, Na-ion transference number ($t_{Na+}$), proving the information of Na-ion mobility and diffusivity is a key parameter for QSS electrolytes [19, 49]. It can be found that (**Fig. 6c and 6d**) the $t_{Na+}$ of the Celgard, PVDF-HFP, and hybrid QSS electrolytes are 0.18, 0.37, and 0.55, respectively. Higher $t_{Na+}$ of the hybrid QSS electrolytes increasing with an increase of UiO-66 loading suggests the ability of activated UiO-66 that facilitates the Na-ion transport. Inactivated UiO-66 based QSS electrolyte (**Fig. S9**) was also investigated and shows dramatically reduced $t_{Na+}$ (0.40), further proving that the open metal sites of activated UiO-66 can capture $PF_6^-$, and thus liberate $Na^+$ for easier migration [19, 70]. This is essential for reducing the kinetics imbalance between anodes and cathodes.

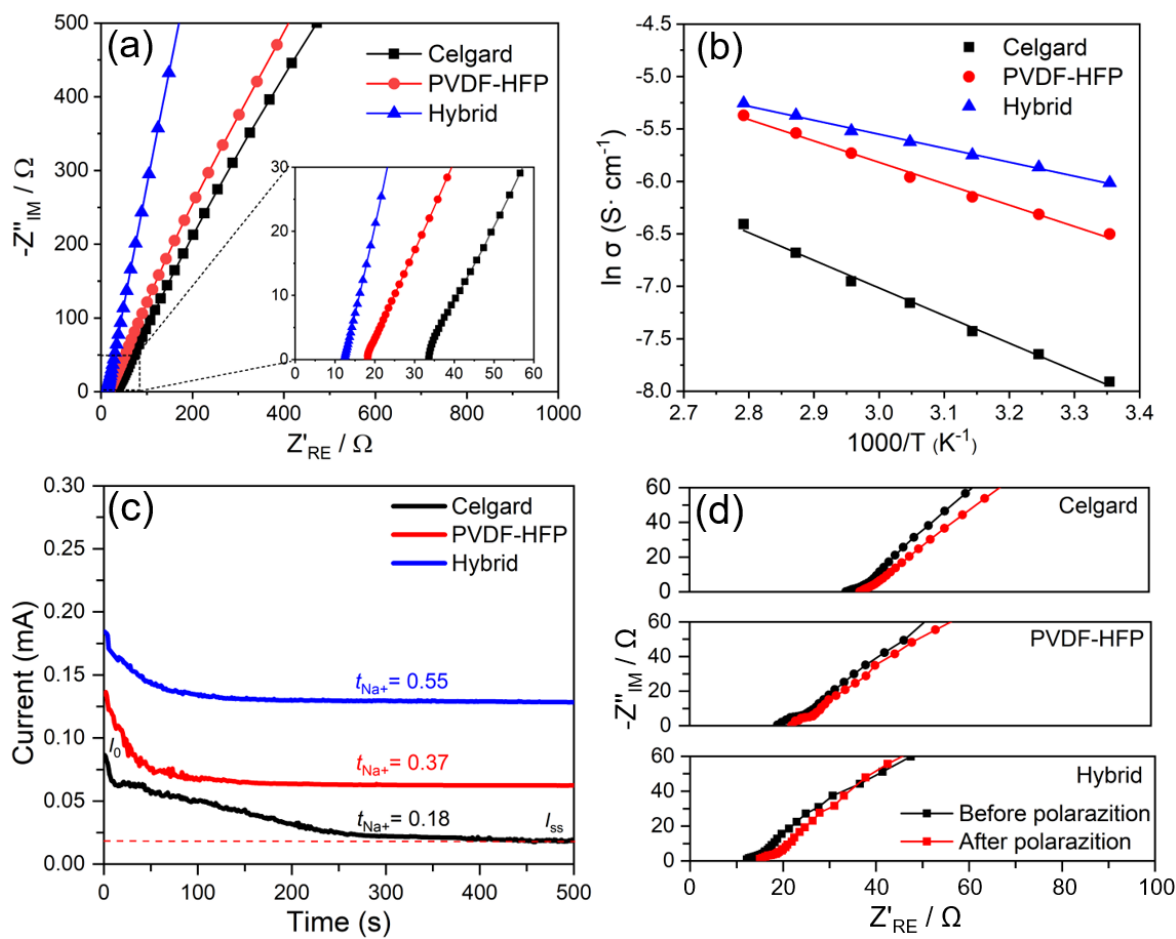

**Fig. 6.** (a) Nyquist plots, (b) Arrhenius plots, (c) Chronoamperometry curves, and (d) Initial and steady-state Nyquist plots of the Celgard, PVDF-HFP, and hybrid separators.

Electrochemical performance of the hybrid QSS electrolytes was initially investigated in Na-ion half-cells to determine the optimal loading of UiO-66 (**Fig. S10a**). The half-cell based on 6-Hybrid electrolyte exhibits superior rate performance compared to 3-Hybrid and 9-Hybrid electrolytes, especially at a higher discharge rate. Inferior rate-performances of hybrid electrolytes containing lower (3 wt%) and higher (9 wt%) UiO-66 loadings are due to the reduced ionic conductivity and non-uniform Na-ion diffusion, respectively, which can seriously affect the fast charge-discharge capability of Na-ion based devices. The capacity loss in the first discharge cycle is related to the SEI formation and electrolyte decomposition (side reactions) on the surface of the anode during the Na-ion intercalation process [7, 11]. It could be minimized through presodiation technique to supply additional Na-ions for the fabrication of full-cells. The hybrid QSS electrolyte (**Fig. 7a**) delivers a high reversible specific capacity and excellent rate-performance (230 mAh·g$^{-1}$ @25 mA·g$^{-1}$ and 70 mAh·g$^{-1}$ @1000 mA·g$^{-1}$) after the first cycle. In comparison, the PVDF-HFP and Celgard show much lower specific capacities and inferior rate-performances. It can be attributed to their lower ionic-conductivity and higher charge-transfer resistance compared to the hybrid electrolyte. This is particularly pronounced during the fast charge-discharge process, such as the nearly disappeared capacity of the Celgard observed under higher current densities. The inactivated UiO-66 based electrolyte (**Fig. S10b**) presents an inferior rate-performance compared to the one based on activated UiO-66 due to the deficient open pores and metal sites that support faster Na-ion transport. In addition, the trace water and residual reactants in the inactivated UiO-66 structure could induce decomposition reactions of the electrolyte solution, e.g. promoting the NaPF$_6$ decompose to more insoluble POF$_3$ and NaF: (a) NaPF$_6$ → NaF + PF$_5$, (b) PF$_5$ + H$_2$O → POF$_3$ + NaF, and deposit on the SEI surface to increase the Na-ion transfer resistance [71]. Such an excessive Na-ion consumption could also increasingly lead to a capacity fading. The hybrid electrolyte shows no plateaus on the sloping voltage profiles (**Fig. 7b**) indicating a dominant pseudocapacitive Na-ion storage process. The CV curves of hybrid electrolyte (**Fig. 7c**) demonstrate identical shapes at various scan rates and overlap during the consecutive cycles, which differs from that of PVDF-HFP and Celgard (**Fig. S11a, and S11b**), implying highly reversible and pseudocapacitive Na-ion storage. The power-law relationship between the peak current density ($i$) and the scan rate ($v$): $i_p = av^b$ has been well established to describe the charge storage mechanism. $b = 0.5$ represents a diffusion-controlled process, and $b = 1.0$ indicates a capacitive-dominated behavior [18]. This can be obtained from the slope of the log($v$)–log($i$) plots. The higher *b* values (**Fig. 7d, Fig. S11c, and S11d**) for anodic/cathodic peaks of the hybrid electrolyte (0.76/0.73) compared to that of the PVDF-HFP (0.64/0.67) and Celgard (0.58/0.59) suggest faster kinetics dominated by a pseudocapacitive process. Pseudocapacitive contributions of the hybrid, PVDF-HFP, and Celgard electrolytes are 76%, 57%, and 36% at a scan rate of 1.0 mV·s$^{-1}$, respectively (**Fig. 7e, Fig. S10e, and S10f**) and increase with an increase of scan rates (**Fig. 7f, Fig. S11g, and S11h**). Such highly pseudocapacitive performance of hybrid electrolyte is the key for SHCs to achieve fast charging-discharging capability, high specific capacity, and superior cycling stability. As expected, the Na-ion half-cell based on hybrid electrolyte (**Fig. 7g**) demonstrates superior capacity retention (90 mAh·g$^{-1}$ @1000 mA·g$^{-1}$ after 100 cycles) compared to PVDF-HFP and Celgard under identical testing conditions. The Nyquist plot (**Fig. S12a**) of the separator demonstrates a nearly unchanged charge transfer resistance before and after cycling indicating that the channels of the separator for Na-ion transport remain unobstructed. Additional peaks are not observed from the XRD pattern (**Fig. S12b**) of the cycled separator indicating the lack of deposited byproducts and excellent chemical stability of UiO-66/PVDF-HFP separator. SEM image (**Fig. S12c**) also presents a stable porous structure due to the good compatibility of UiO-66/PVDF-HPF separator with the organic electrolyte solution. The Nyquist plots of Na-ion half-cells based on these electrolytes (**Fig. 7h**) exhibit a semicircle in the high-medium-frequency region indicating the overall resistance including the equivalent series resistance ($R_s$) and charge-transfer resistance ($R_{ct}$), and a sloping line in the low-frequency region related to Na-ion diffusion resistance ($Z_w$) within the electrodes [7]. The $R_{ct}$ of the hybrid electrolyte (123 Ω) is much lower than that of the PVDF-HFP (172 Ω) and Celgard (291 Ω), suggesting an improved charge transfer kinetics. Na-ion diffusion coefficients calculated for the hybrid, PVDF-HFP, and Celgard electrolytes are 1.8×10$^{-14}$, 1.1×10$^{-14}$ and 2.3×10$^{-15}$ cm$^2$·s$^{-1}$, respectively. The rapid Na-ion diffusion of hybrid electrolyte into electrodes is a result of good electrolyte/electrode affinity, fast Na-ion transport, and the high pseudocapacitance of the dual-phase $TiO_2$ nanosheet electrode reported earlier [7].

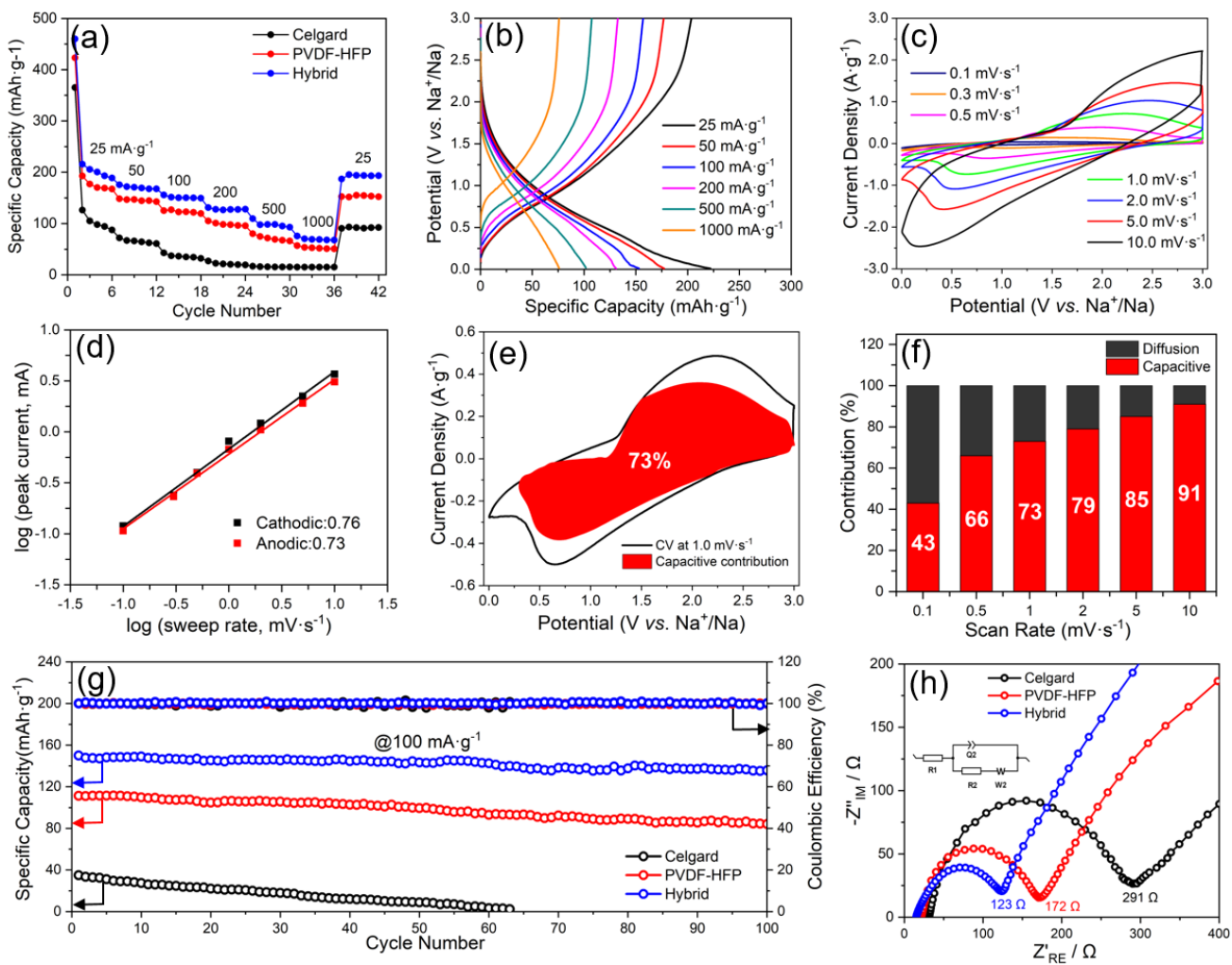

**Fig. 7**. (a) Rate performance of quasi-solid-state Na-ion half-cells based on the Celgard, PVDF-HFP, and hybrid separators. (b) Charge-discharge profiles, (c) CV curves, (d) Scan-rate dependence of peak current (e) Pseudocapacitive contribution at 1.0 mV·s$^{-1}$, (f) Capacitive contribution percentages at various scan rates of the hybrid separator. (g) cycling performance and (h) Nyquist plots of the Celgard, PVDF-HFP, and hybrid separators.

A hybrid separator was finally employed to fabricate quasi-solid-state Na-ion hybrid capacitor (QSS-SHC) assembled with dual-phase $TiO_2$ anode and AC cathode (**Fig. 8a**). During the charging process, $PF_6^-$ anions in the electrolyte migrate to the cathode and adsorb onto the surface *via* a fast "non-Faradaic reaction". Meanwhile, Na-ions migrate to the anode and store *via* a sluggish "Faradaic reaction". Generally, the kinetics-mismatch of these different storage mechanisms could result in a serious concentration polarization. In the case of using the hybrid electrolyte, the $PF_6^-$ anions can be partially captured by the open metal sites of the activated UiO-66, and more Na-ions are thus liberated to migrate dramatically faster to the anode. Such reduced effect of concentration polarization on charge separation is beneficial for achieving higher energy and power densities as well as better cycling stability of the QSS-SHC. To investigate the feasibility of the QSS-SHC, $TiO_2$ anode is integrated with an activated carbon (AC) cathode. CV curves of anode, cathode, and QSS-SHC in the voltage ranges of 0−3.0, 1.0−4.0, and 1.0−4.0 V, respectively are presented in **Fig. S13**. Square type CV curve of AC cathode indicates the typical double-layer type charge storage mechanism. Operating voltage window of the QSS-SHC is limited to 1.0–4.0 V to avoid electrolyte decomposition, ensure the stability of AC cathode, and take full advantage of the pseudocapacitive Na-ion storage. The near-rectangular CV curve of QSS-SHC indicates the charge storage contribution from the synergy between pseudocapacitive reaction of the anode and non-Faradaic double-layer process of the cathode. The QSS-SHC (**Fig. 8b**) exhibits an excellent rate-performance with a discharge capacity of 202 mAh·g$^{-1}$ @25 mA·g$^{-1}$ and even 50 mAh·g$^{-1}$ @1000 mA·g$^{-1}$. The charge-discharge voltage-profiles (**Fig. 8c**) and CV curves (**Fig. 8d**) at various current densities present a near-linear correlation of capacitive properties, indicating rapid Na-ion storage kinetics. The QSS-SHC demonstrates ultralong cycle life of 10000 cycles @1000 mA·g$^{-1}$ with a high capacity-retention of ~ 80 % and coulombic efficiency of ~ 100 % (**Fig. 8e**). Furthermore, the maximum energy of 182 Wh·kg$^{-1}$ @31 W·kg$^{-1}$ and power density of 5280 W·kg$^{-1}$ @22 Wh·kg$^{-1}$ are very close to that of the liquid-state SHC (**Fig. S14**) and exceptional among the previous reports of SHCs (**Fig. 8f**) [12-18]. This result demonstrates that the UiO-66/PVDF-HFP hybrid separator can facilitate the Na-ion transport from electrolyte to electrode and minimize the kinetics mismatch between cathode and anode.

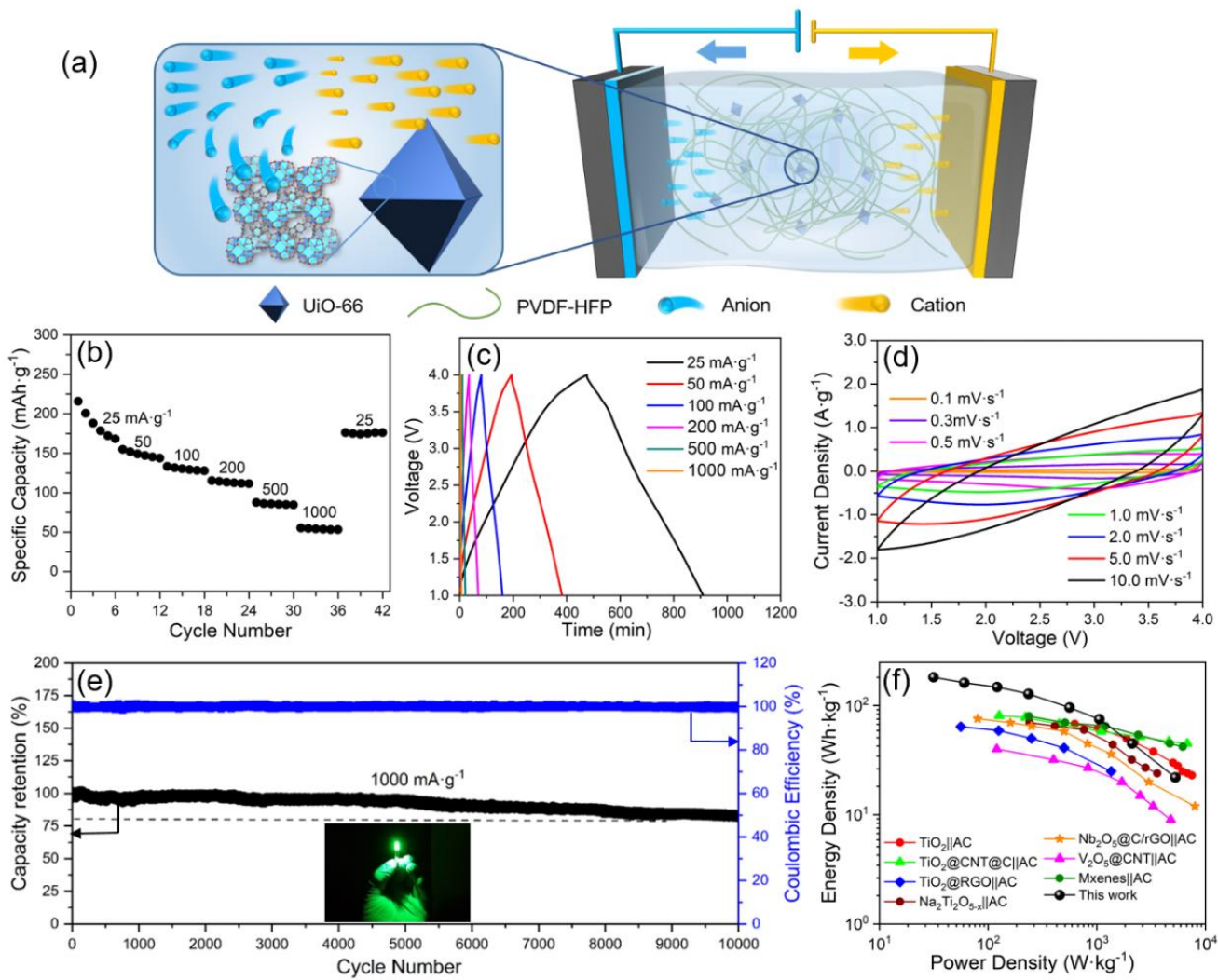

**Fig. 8**. (a) Working mechanism, (b) Rate performance, (c) Charge-discharge profiles at corresponding current densities, (d) CV curves at various scan rates, (e) Cycling performance after rate performance, and (f) Ragone plot of the quasi-solid-state hybrid capacitor.

**4. Conclusions.**

In summary, activated UiO-66 MOF was prepared and employed as a multifunctional filler to modify the porous PVDF-HFP separator. The hybrid separator demonstrates high porosity (> 70%) and electrolyte uptake (~ 320%), which are essential for the fabrication of quasi-solid-state electrolyte with sufficient Na-ion supply and good electrode/electrolyte affinity. Moreover, the good mechanical strength and dimensional thermal stability of the hybrid separator can ensure a safe SHC operation by preventing the electrode short circuit related to the separator deformation. The hybrid separator exhibits a huge reduction (75%) in pHRR from MCC, indicating a significantly improved flame retardancy property. This is resulted from the transformation of UiO-66 into $ZrO_2$ accompanied by the consumption of oxygen and the formation of the barrier char that suppress the further heat release. High ionic conductivity (2.44 mS·cm$^{-1}$) and Na-ion transference number (0.55) of quasi-solid-state reveal that the open metal sites of UiO-66 can capture the $PF_6^-$ in electrolyte and liberate the $Na^+$ for faster migration during discharging process. The low MacMullin number (3.2) and activation energy (1.33 kJ·mol$^{-1}$) proves that reduced crystallinity and good wettability of the hybrid separator are beneficial for achieving a high Na-ions permeability. Highly pseudocapacitive Na-ion storage is realized based on the premise of the improved charge separation and charge transfer. The quasi-solid-state Na-ion hybrid capacitor based on this separator exhibits high energy density (182 Wh·kg$^{-1}$ @31 W·kg$^{-1}$) and power density (5280 W·kg$^{-1}$ @22 Wh·kg$^{-1}$), as well as excellent cycling stability (10000 cycles @1000 mA·g$^{-1}$). Hence, the demonstrated multifunctional separator has great potential for the application in Na-ion hybrid capacitors with both improved fire safety and electrochemical performance.

**Declaration of competing interest**

The authors declare that they have no known competing financial interests or personal relationships that could have appeared to influence the work reported in this paper.

**Acknowledgments**

We acknowledge IMDEA Materials Institute STRUBAT Project, Spanish Ministry of Economy, Industry, and Competitiveness (MINECO), Spanish Ministry of Science and Innovation, and Comunidad de Madrid for Juan de la Cierva fellowship (IJCI-2015-25488), Retos investigacion project (MAT2017-84002-C2-2-R) / Ramon y Cajal fellowship (RYC-2018-025893-I), Talent attraction fellowship (2016-T1/IND-1300), and China Scholarship Council (No. 201706740087) for financial support.

**Appendix A. Supplementary data**

Supplementary data to this article can be found online.

**Reference**

[1] K. Liu, Y. Liu, D. Lin, A. Pei, Y. Cui, Materials for Lithium-Ion Battery Safety, Sci. Adv., 4 (2018), 9820.

[2] P.K. Nayak, L. Yang, W. Brehm, P. Adelhelm, From Lithium-Ion to Sodium-Ion Batteries: Advantages, Challenges, and Surprises, Angew. Chem. Int. Ed., 57 (2018), 102−120.

[3] B. Babu, P. Simon, A. Balducci, Fast Charging Materials for High Power Applications, Adv. Energy Mater., 10 (2020), 2001128.

[4] W. Zuo, R. Li, C. Zhou, Y. Li, J. Xia, J. Liu, Battery-Supercapacitor Hybrid Devices: Recent Progress and Future Prospects, Adv. Sci., 4 (2017), 1600539.

[5] Y. Zhang, J. Jiang, Y. An, L. Wu, H. Dou, J. Zhang, Y. Zhang, S. Wu, M. Dong, X. Zhang, Z. Guo, Sodium-ion capacitors: Materials, Mechanism, and Challenges, ChemSusChem, 13 (2020), 2522−2539.

[6] Y. Shao, M.F. El-Kady, J. Sun, Y. Li, Q. Zhang, M. Zhu, H. Wang, B. Dunn, R.B. Kaner, Design and Mechanisms of Asymmetric Supercapacitors, Chem. Rev., 118 (2018), 9233–9280.

[7] W. Feng, R.R. Maca, V. Etacheri, High-Energy-Density Sodium-Ion Hybrid Capacitors Enabled by Interface-Engineered Hierarchical $TiO_2$ Nanosheet Anodes, ACS Appl. Mater. Interfaces, 12 (2020), 4443–4453.

[8] F. Wang, X. Wang, Z. Chang, X. Wu, X. Liu, L. Fu, Y. Zhu, Y. Wu, W. Huang, A Quasi-Solid-State Sodium-Ion Capacitor with High Energy Density, Adv. Mater., 27 (2015), 6962−6968.

[9] J. Ding, H. Wang, Z. Li, K. Cui, D. Karpuzov, X. Tan, A. Kohandehghan, D. Mitlin, Peanut Shell Hybrid Sodium Ion Capacitor with Extreme Energy–Power Rivals Lithium Ion Capacitors, Energy Environ. Sci., 8 (2015), 941−955.

[10] Z. Le, F. Liu, P. Nie, X. Li, X. Liu, Z. Bian, G. Chen, H.B. Wu, Y. Lu, Pseudocapacitive Sodium Storage in Mesoporous Single-Crystal-like $TiO_2$-Graphene Nanocomposite Enables High-Performance Sodium-Ion Capacitors, ACS Nano, 11 (2017), 2952−2960.

[11] N. Kurra, M. Alhabeb, K. Maleski, C.-H. Wang, H.N. Alshareef, Y. Gogotsi, Bistacked Titanium Carbide (MXene) Anodes for Hybrid Sodium-Ion Capacitors, ACS Energy Lett., 3 (2018), 2094−2100.

[12] Y.-E. Zhu, L. Yang, J. Sheng, Y. Chen, H. Gu, J. Wei, Z. Zhou, Fast Sodium Storage in $TiO_2$@CNT@C Nanorods for High-Performance Na-Ion Capacitors, Adv. Energy Mater., 7 (2017), 1701222.

[13] B. Babu, S.G. Ullattil, R. Prasannachandran, J. Kavil, P. Periyat, M.M. Shaijumon, $Ti^{3+}$ Induced Brown $TiO_2$ Nanotubes for High Performance Sodium-Ion Hybrid Capacitors, ACS Sustainable Chem. Eng., 6 (2018), 5401–5412.

[14] H. Huang, D. Kundu, R. Yan, E. Tervoort, X. Chen, L. Pan, M. Oschatz, M. Antonietti, M. Niederberger, Fast Na-Ion Intercalation in Zinc Vanadate for High-Performance Na-Ion Hybrid Capacitor, Adv. Energy Mater., 8 (2018), 1802800.

[15] L.-F. Que, F.-D. Yu, K.-W. He, Z.-B. Wang, D.-M. Gu, Robust and Conductive $Na_2Ti_2O_{5-x}$ Nanowire Arrays for High-Performance Flexible Sodium-Ion Capacitor, Chem. Mater., 29 (2017), 9133–9141.

[16] Z. Chen, V. Augustyn, X. Jia, Q. Xiao, B. Dunn, Y. Lu, High-performance sodium-ion pseudocapacitors based on hierarchically porous nanowire composites, ACS Nano, 6 (2012), 4319−4327.

[17] E. Lim, C. Jo, M.S. Kim, M.-H. Kim, J. Chun, H. Kim, J. Park, K.C. Roh, K. Kang, S. Yoon, J. Lee, High-Performance Sodium-Ion Hybrid Supercapacitor Based on Nb2O5@Carbon Core-Shell Nanoparticles and Reduced Graphene Oxide Nanocomposites, Adv. Funct. Mater., 26 (2016), 3711–3719.

[18] J. Luo, C. Fang, C. Jin, H. Yuan, O. Sheng, R. Fang, W. Zhang, H. Huang, Y. Gan, Y. Xia, C. Liang, J. Zhang, W. Li, X. Tao, Tunable pseudocapacitance storage of MXene by cation pillaring for high performance sodium-ion capacitors, J. Mater. Chem. A, 6 (2018), 7794–7806.

[19] C.-C. Sun, A. Yusuf, S.-W. Li, X.-L. Qi, Y. Ma, D.-Y. Wang, Metal Organic Frameworks Enabled Rational Design of Multifunctional PEO-based Solid Polymer Electrolytes, Chem. Eng. J., 414 (2021), 128702.

[20] S. Chen, J. Zheng, L. Yu, X. Ren, M.H. Engelhard, C. Niu, H. Lee, W. Xu, J. Xiao, J. Liu, J.-G. Zhang, High-Efficiency Lithium Metal Batteries with Fire-Retardant Electrolytes, Joule, 2 (2018), 1548−1558.

[21] J. Yang, M. Zhang, Z. Chen, X. Du, S. Huang, B. Tang, T. Dong, H. Wu, Z. Yu, J. Zhang, G. Cui, Flame-Retardant Quasi-Solid Polymer Electrolyte Enabling Sodium Metal Batteries with Highly Safe Characteristic and Superior Cycling Stability, Nano Research, 12 (2019), 2230−2237.

[22] Y. Yu, H. Che, X. Yang, Y. Deng, L. Li, Z.-F. Ma, Non-Flammable Organic Eectrolyte for Sodium-Ion Batteries, Electrochem. Commun., 110 (2020), 106635.

[23] G. Zeng, J. Zhao, C. Feng, D. Chen, Y. Meng, B. Boateng, N. Lu, W. He, Flame-Retardant Bilayer Separator with Multifaceted van der Waals Interaction for Lithium-Ion Batteries, ACS Appl. Mater. Interfaces, 11 (2019), 26402−26411.

[24] J.W. Lee, A.M. Soomro, M. Waqas, M.A.U. Khalid, K.H. Choi, A Highly Efficient Surface Modified Separator Fabricated with Atmospheric Atomic Layer Deposition for High Temperature Lithium Ion Batteries, Int. J. Energy Res., 44 (2020), 7035−7046.

[25] J.K. Pi, G.P. Wu, H.C. Yang, C.G. Arges, Z.K. Xu, Separators with Biomineralized Zirconia Coatings for Enhanced Thermo- and Electro-Performance of Lithium-Ion Batteries, ACS Appl. Mater. Interfaces, 9 (2017), 21971−721978.

[26] H. Han, S. Niu, Y. Zhao, T. Tan, Y. Zhang, $TiO_2$/Porous Carbon Composite-Decorated Separators for Lithium/Sulfur Battery, Nanoscale Res. Lett., 14 (2019), 176.

[27] Q.S. Rao, S.Y. Liao, X.W. Huang, Y.Z. Li, Y.D. Liu, Y.G. Min, Assembly of MXene/PP Separator and Its Enhancement for Ni-Rich $LiNi_{0.8}Co_{0.1}Mn_{0.1}O_2$ Electrochemical Performance, Polymers, 12 (2020), 2192.

[28] Z. Zhang, Y. Lai, Z. Zhang, K. Zhang, J. Li, $Al_2O_3$-Coated Porous Separator for Enhanced Electrochemical Performance of Lithium Sulfur Batteries, Electrochim. Acta, 129 (2014), 55−61.

[29] S. Manoharan, P. Pazhamalai, V.K. Mariappan, K. Murugesan, S. Subramanian, K. Krishnamoorthy, S.-J. Kim, Proton Conducting Solid Electrolyte-Piezoelectric PVDF Hybrids: Novel Bifunctional Separator for Self-Charging Supercapacitor Power Cell, Nano Energy, 83 (2021), 105753.

[30] K. Krishnamoorthy, P. Pazhamalai, V.K. Mariappan, S.S. Nardekar, S. Sahoo, S.J. Kim, Probing the Energy Conversion Process in Piezoelectric-Driven Electrochemical Self-Charging Supercapacitor Power Cell Using Piezoelectrochemical Spectroscopy, Nat. Commun., 11 (2020), 2351.

[31] X. Xue, P. Deng, S. Yuan, Y. Nie, B. He, L. Xing, Y. Zhang, CuO/PVDF Nanocomposite Anode for A Piezo-Driven Self-Charging Lithium Battery, Energy Environ. Sci., 6 (2013), 2615–2620.

[32] H. He, Y. Fu, T. Zhao, X. Gao, L. Xing, Y. Zhang, X. Xue, All-Solid-State Flexible Self-Charging Power Cell basing on Piezo-Electrolyte for Harvesting/Storing Body-Motion Energy and Powering Wearable Electronics, Nano Energy, 39 (2017), 590–600.

[33] F.I. Saaid, T.-Y. Tseng, T. Winie, PVdF-HFP Quasi-Solid-State Electrolyte for Application in Dye-Sensitized Solar Cells, Int. J. Technol, 9 (2018), 1187−1195.

[34] K. Liu, W. Liu, Y. Qiu, B. Kong, Y. Sun, Z. Chen, D. Zhuo, D. Lin, Y. Cui, Electrospun Core-Shell Microfiber Separator with Thermal Triggered Flame Retardant Properties for Lithium Ion Battery, Sci. Adv., 3 (2017), e1601978.

[35] S. Janakiraman, A. Surendran, S. Ghosh, S. Anandhan, A. Venimadhav, Electroactive Poly(Vinylidene Fluoride) Fluoride Separator for Sodium Ion Battery with High Coulombic Efficiency, Solid State Ionics, 292 (2016), 130−135.

[36] A. Valverde, R. Gonçalves, M.M. Silva, S. Wuttke, A. Fidalgo-Marijuan, C.M. Costa, J.L. Vilas-Vilela, J.M. Laza, M.I. Arriortua, S. Lanceros-Méndez, R. Fernández de Luis, Metal-Organic Framework Based PVDF Separators for High Rate Cycling Lithium-Ion Batteries, ACS Appl. Energy Mater., 3 (2020), 11907−11919.

[37] L. Li, H. Li, Y. Wang, S. Zheng, Y. Zou, Z. Ma, Poly(Vinylidenefluoride-Hexafluoropropylene)/ Cellulose/Carboxylic TiO2 Composite Separator with High Temperature Resistance for Lithium-Ion Batteries, Ionics, 26 (2020), 4489−4497.

[38] D. Wu, L. Deng, Y. Sun, K.S. Teh, C. Shi, Q. Tan, J. Zhao, D. Sun, L. Lin, A High-Safety PVDF/Al2O3 Composite Separator for Li-Ion Batteries via Tip-Induced Electrospinning and Dip-Coating, RSC Adv., 7 (2017), 24410−24416.

[39] A.K. Solarajan, V. Murugadoss, S. Angaiah, High Performance Electrospun PVdF-HFP/SiO2 Nanocomposite Membrane Electrolyte for Li-Ion Capacitors, J. Appl. Polym. Sci., 134 (2017), 45177.

[40] H. Zhang, J. Hou, Y. Hu, P. Wang, R. Ou, L. Jiang, J.Z. Liu, B.D. Freeman, A.J. Hill, H. Wang, Ultrafast Selective Transport of Alkali Metal Ions in Metal Organic Frameworks with Subnanometer Pores, Sci. Adv., 4 (2018), 0066.

[41] T. Xu, M.A. Shehzad, X. Wang, B. Wu, L. Ge, T. Xu, Engineering Leaf-Like UiO-66-SO3H Membranes for Selective Transport of Cations, Nano-Micro Lett., 12 (2020), 51.

[42] J. Luo, Y. Li, H. Zhang, A. Wang, W.S. Lo, Q. Dong, N. Wong, C. Povinelli, Y. Shao, S. Chereddy, S. Wunder, U. Mohanty, C.K. Tsung, D. Wang, A Metal-Organic Framework Thin Film for Selective Mg2+ Transport, Angew. Chem. Int. Ed., 58 (2019), 15313−15317.

[43] C. Zhang, L. Shen, J. Shen, F. Liu, G. Chen, R. Tao, S. Ma, Y. Peng, Y. Lu, Anion-Sorbent Composite Separators for High-Rate Lithium-Ion Batteries, Adv. Mater., 31 (2019), 1808338.

[44] J. Zhang, Z. Li, X.-L. Qi, D.-Y. Wang, Recent Progress on Metal–Organic Framework and Its Derivatives as Novel Fire Retardants to Polymeric Materials, Nano-Micro Lett., 12 (2020), 173.

[45] T. Sai, S. Ran, Z. Guo, Z. Fang, A Zr-based Metal Organic Frameworks Towards Improving Fire Safety and Thermal Stability of Polycarbonate, Compos. B Eng., 176 (2019), 107198.

[46] J. Ran, J. Qiu, H. Xie, X. Lai, H. Li, X. Zeng, Combination Effect of Zirconium Phosphate Nanosheet and PU-coated Carbon Fiber on Flame Retardancy and Thermal Behavior of PA46/PPO Alloy, Compos. B. Eng., 166 (2019), 621–632.

[47] X.-Q. Liu, D.-Y. Wang, X.-L. Wang, L. Chen, Y.-Z. Wang, Synthesis of Organo-Modified α-Zirconium Phosphate and Its Effect on the Flame Retardancy of IFR Poly(Lactic Acid) Systems, Polym. Degrad. Stab., 96 (2011), 771−777.

[48] M.J. Katz, Z.J. Brown, Y.J. Colon, P.W. Siu, K.A. Scheidt, R.Q. Snurr, J.T. Hupp, O.K. Farha, A Facile Synthesis of UiO-66, UiO-67 and Their Derivatives, Chem. Commun., 49 (2013), 9449–9451.

[49] H. Yang, B. Liu, J. Bright, S. Kasani, J. Yang, X. Zhang, N. Wu, A Single-Ion Conducting UiO-66 Metal–Organic Framework Electrolyte for All-Solid-State Lithium Batteries, ACS Appl. Energy Mater., 3 (2020), 4007−4013.

[50] J. Zhang, Z. Li, Z.-B. Shao, L. Zhang, D.-Y. Wang, Hierarchically Tailored Hybrids via Interfacial-Engineering of Self-Assembled UiO-66 and Prussian Blue Analogue: Novel Strategy to Impart Epoxy High-Efficient Fire Retardancy and Smoke Suppression, Chem. Eng. J., 400 (2020), 125942.

[51] C. Ribeiro, C.M. Costa, D.M. Correia, J. Nunes-Pereira, J. Oliveira, P. Martins, R. Goncalves, V.F. Cardoso, S. Lanceros-Mendez, Electroactive Poly(Vinylidene Fluoride)-based Structures for Advanced Applications, Nat Protoc, 13 (2018), 681−704.

[52] V.F. Cardoso, G. Botelho, S. Lanceros-Méndez, Nonsolvent Induced Phase Separation Preparation of Poly(vinylidene Fluoride-co-Chlorotrifluoroethylene) Membranes with Tailored Morphology, Piezoelectric Phase Content and Mechanical Properties, Mater. Des., 88 (2015), 390−397.

[53] G. Xu, P. Nie, H. Dou, B. Ding, L. Li, X. Zhang, Exploring Metal Organic Frameworks for Energy Storage in Batteries and Supercapacitors, Mater. Today, 20 (2017), 191−209.

[54] M.H. Zeng, X.L. Feng, X.M. Chen, Crystal-to-Crystal Transformations of A Microporous Metal-Organic Laminated Framework Triggered by Guest Exchange, Dehydration and Readsorption, Dalton Trans, (2004), 2217−2223.

[55] Q. Yang, H.Y. Zhang, L. Wang, Y. Zhang, J. Zhao, Ru/UiO-66 Catalyst for the Reduction of Nitroarenes and Tandem Reaction of Alcohol Oxidation/Knoevenagel Condensation, ACS Omega, 3 (2018), 4199−4212.

[56] F. Liu, N.A. Hashim, Y. Liu, M.R.M. Abed, K. Li, Progress in the production and modification of PVDF membranes, J. Membr. Sci., 375 (2011), 1−27.

[57] G.R. Guillen, Y. Pan, M. Li, E.M.V. Hoek, Preparation and Characterization of Membranes Formed by Nonsolvent Induced Phase Separation: A Review, Ind. Eng. Chem. Res., 50 (2011), 3798−3817.

[58] J. Cannarella, X. Liu, C.Z. Leng, P.D. Sinko, G.Y. Gor, C.B. Arnold, Mechanical Properties of a Battery Separator under Compression and Tension, J. Electrochem. Soc., 161 (2014), F3117−F3122.

[59] X. Cai, T. Lei, D. Sun, L. Lin, A Critical Analysis of the α, β and γ Phases in Poly(Vinylidene Fluoride) using FTIR, RSC Adv., 7 (2017), 15382−15389.

[60] J.M. Yang, H.Z. Wang, C.C. Yang, Modification and Characterization of Semi-Crystalline Poly(Vinyl Alcohol) with Interpenetrating Poly(Acrylic Acid) by UV Radiation Method for Alkaline Solid Polymer Electrolytes Membrane, J. Membr. Sci., 322 (2008), 74−80.

[61] M. Hirschler, Effect of Oxygen on the Thermal Decomposition of Poly (Vinylidene Fluoride), Eur. Polym. J., 18 (1982), 463−467.

[62] Y. Gao, J. Wu, Q. Wang, C.A. Wilkie, D. O'Hare, Flame retardant polymer/layered double hydroxide nanocomposites, J. Mater. Chem. A, 2 (2014), 10996−11016.

[63] Y. Kim, S. Lee, H. Yoon, Fire-Safe Polymer Composites: Flame-Retardant Effect of Nanofillers, Polymers, 13 (2021), 540.

[64] L. Wang, Z. Wang, Y. Sun, X. Liang, H. Xiang, Sb2O3 Modified PVDF-CTFE Electrospun Fibrous Membrane as A Safe Lithium-Ion Battery Separator, J. Membr. Sci., 572 (2019), 512−519.

[65] J.Y. Xu, J. Liu, K.D. Li, L. Miao, S. Tanemura, Novel PEPA-Functionalized Graphene Oxide for Fire Safety Enhancement of Polypropylene, Sci. Technol. Adv. Mater., 16 (2015), 025006.

[66] F. Zhang, J. Zhang, C. Jiao, Study on Char Structure of Intumescent Flame-Retardant Polypropylene, Polym. Plast. Technol. Eng., 47 (2008), 1179−1186.

[67] Narayan Ramesh, J.L. Duda, A Modified Free-Volume Model: Correlation of Ion-Conduction in Strongly Associating Polymeric Materials, J. Membr. Sci., 191 (2001), 13–30.

[68] Y.Q. Yang, Z. Chang, M.X. Li, X.W. Wang, Y.P. Wu, A Sodium Ion Conducting Gel Polymer Electrolyte, Solid State Ionics, 269 (2015), 1–7.

[69] K. Prasanna, C.W. Lee, Physical, Thermal, and Electrochemical Characterization of Stretched Polyethylene Separators for Application in Lithium-Ion Batteries, J. Solid State Electrochem., 17 (2013), 1377–1382.

[70] H. Chen, Y. Xiao, C. Chen, J. Yang, C. Gao, Y. Chen, J. Wu, Y. Shen, W. Zhang, S. Li, F. Huo, B. Zheng, Conductive MOF-Modified Separator for Mitigating the Shuttle Effect of Lithium-Sulfur Battery through a Filtration Method, ACS Appl. Mater. Interfaces, 11 (2019), 11459–11465.

[71] X. Cui, F. Tang, Y. Zhang, C. Li, D. Zhao, F. Zhou, S. Li, H. Feng, Influences of Trace Water on Electrochemical Performances for Lithium Hexafluoro Phosphate- and Lithium Bis(oxalato)Borate-based Electrolytes, Electrochim. Acta, 273 (2018), 191–199.